\documentclass[preprint,twocolumn]{revtex4-2}

\usepackage{xcolor}
\usepackage{placeins}
\usepackage{graphicx}

\begin{document}

\title{Massive Enhanced Extracted Email Features Tailored for Cosine Distance}
\author{Farshad Barahimi}
\email{farshad.barahimi@dal.ca}
\affiliation{Dalhousie University, Canada\\{\small(Farshad Barahimi is a PhD student at Dalhousie University)}}

\begin{abstract}
In this paper, the process of converting the Enron email dataset~\cite{1_enron_email_dataset} to thousands of features per email for a selected set of 2400 labelled emails is explained and evaluated. The final features are tailored for Cosine distance so that the Cosine distance invertly reflect the number of top indicative words of each email that are common between the two emails in an explainable normalized fashion. The labelling is based on the leaf folder name in the Enron email dataset~\cite{1_enron_email_dataset} folders tree and the 2400 emails selected consist 300 emails for each of the 8 labels. The evaluation is based on the accuracy of a $k$ nearest neighbours majority voting classification using Cosine distance. In addition to KNN majority voting classification accuracy and confusion matrix, some statistics for the process is reported. The KNN majority voting classification accuracy using Cosine distance is 76.75\% which shows at least some level of success given the 8 labels involved. The result of conversion is 48557 features per selected email out of which exactly 40 features per email are non-zero. The result of conversion is a data set named MeeefTCD (Massive Enhanced Extracted Email Features Tailored for Cosine Distance) available at \url{https://web.cs.dal.ca/~barahimi/data-sets/meeeftcd/} and on a github repository mentioned in this paper.
\end{abstract}

\maketitle

\section{Introduction}
Extracting text document features~\cite{4_feature_extraction,5_review,6_mining_opinions,7_extracting_knowledge} has applications such as text document classification~\cite{2_text_classification,5_review} and visualization (embedding in two dimensions) of text documents collections~\cite{8_embedding}. While many techniques exist for text classification and embedding of data features, better benchmark data sets of text features can help evaluation of text classification techniques or data features embedding techniques usable on text documents collections or any classification technique usable on text features.

The Enron email dataset~\cite{1_enron_email_dataset}, is a publicly available data set of Enron emails which was originally prior to modifications made public and posted online by the United States Federal Energy Regulatory Commission during an investigation and later a modified version of it was posted online by Cohen~\cite{1_enron_email_dataset}. The modified version labelled May 7, 2015 and posted online by Cohen~\cite{1_enron_email_dataset} is used in this paper to produce email features for a selected set of 2400 emails. The emails in Enron email dataset~\cite{1_enron_email_dataset} are organized in folders which in some cases are reasonable choice for labelling emails.

While different methods exist to measure dissimilarity of text documents in order to better organize the documents, the Cosine distance of Term Frequency–Inverse Document Frequency is a common method but criticized~\cite{2_text_classification,3_beyond}. In short, in a common approach a text document such as an email is seen as a vector of constant number of numbers each relating to the importance of a word of a fixed dictionary for that document. Then to measure distance of two text documents specified by two vectors $V_a=(v_{a_1},v_{a_2},...,v_{a_m})$ and $V_b=(v_{b_1},v_{b_2},...,v_{b_m})$, the Cosine distance of the two vectors denoted by is used which is equal to :
\begin{equation}
1-\frac{\sum\limits_{1 \le i \le m}{(v_{a_i}\cdot v_{b_i})}}{\sqrt{\sum\limits_{1 \le i \le m}{v_{a_i}^2}}\cdot \sqrt{\sum\limits_{1 \le i \le m}{v_{b_i}^2}}}
\label{cosine_distance}
\end{equation}
An important observation which is the key piece of this paper is that if in the formula above (Equation~\ref{cosine_distance}) there is a restriction that the value of each $v_{a_i}$ or $v_{b_i}$ is restricted to be either 1 or 0 (binary), and if the number of ones in $V_a$ is the same as the number of ones in $V_b$, then Cosine distance essentially acts like a digital \textit{and} operator counter which counts the number of places where both vectors have ones normalized by a constant factor. This is important to have a meaningful explanation of the meaning of the Cosine distance. The two other important factors to consider in order to have a meaningful explanation of the meaning of Cosine distance is the meaning of each 1 and the number of 1's in each vector. This paper strives for the meaning of a 1 to be being among top indicative words for that email where definition of top indicative words is as defined in this paper in Section~\ref{features_generation_process} and subsequently since the number of 1's won't be fixed as desired, some secondary 1's are added in extra vector dimensions (increasing m) to have a fixed the number of 1's for each of all vectors. This changes each vector $V_a$ to a vector $\widehat{V}_a=(v_{a_1},v_{a_2},...,v_{a_m},v_{a_{(m+1)}},...v_{a_q})$ where the first $m$ dimensions are called primary dimensions and the next $q$ dimensions are called secondary dimensions. The desired number of 1's per vector is based on some statistical analysis described in Section~\ref{features_generation_process}, however it is important that when a dummy 1 is added to a vector in a secondary dimension, the other vectors don't have a one in that dimension, so it only affects normalization but not counting. While the first $m$ dimensions help counting, the next $q$ dimensions help with consistent constant normalization.

To measure the importance of a dictionary word for a text document, a common method is Term Frequency–Inverse Document Frequency~\cite{2_text_classification,3_beyond} which has some variants but in this paper the variant defined by the following formula is used:
$$v_{a_i}=\frac{F_{a_i}}{\sum_t{F_{a_t}}} \cdot\log(\frac{N}{\widehat{F}_i})$$ where $F_{a_i}$ is the number of times that the $i^{th}$ word of dictionary appears in the email $a$, and $\widehat{F}_i$ specifies the number of emails that contain the the $i^{th}$ word of dictionary at least once, and $N$ is the number of text documents.

Having this introduction, Section~\ref{features_generation_process} goes through the details of generating features for emails that are tailored for Cosine distance. It is important to note that this introduction missed some of important details and steps that are described in Section~\ref{features_generation_process}. In Section~\ref{evaluation_and_statistics} not only accuracy of a KNN majority voting classification is reported but also the confusion matrix is discussed along some statistics. Finally Section~\ref{final_data_set} talks about how the data set is available and also where to find the code used to create the data set and produce the results of this paper.

\onecolumngrid

\section{Features generation process}
\label{features_generation_process}
\FloatBarrier
\subsection{Selecting emails}
The first 300 emails from each of 8 folders labelled projects, logistics, resumes, universities, online\_trading, meetings,management and ces are selected. Due to the nature of the Enron email dataset~\cite{1_enron_email_dataset}, it is possible that multiple folders are labelled the same, in which case only the first one in file traversal of the emails root folder which passes the criteria of having at least 300 emails is chosen. In this paper, the ordered set of selected emails are denoted by $E=(e_1,e_2,...,e_{2400})$. For each of the labels a number between 0 and 7 is used in this paper and the final data set (MeeefTCD) as shown in Table~\ref{label_numbers}.

\begin{table}
\begin{tabular}{| c | c |}
\hline
label & label number \\
\hline
projects & 0 \\
\hline
logistics &1 \\
\hline
resumes & 2 \\
\hline
universities & 3 \\
\hline
online\_trading & 4 \\
\hline
meetings & 5 \\
\hline
management & 6 \\
\hline
ces & 7 \\
\hline
\end{tabular}
\caption{Labels and their corresponding number.}
\label{label_numbers}
\end{table}

\FloatBarrier

\subsection{Lexical parsing}
\subsubsection{Parsing into words}
All of the 2400 emails are parsed into words using space character and any of the characters \textcolor{blue}{$,:!=;'>[]()$} as delimiters after being converted to lowercase text. The lines of emails are trimmed for empty spaces at the beginning and at the end. For each email, except for the subject line, any line above the line starting with \textcolor{blue}{x-filename:} is ignored to ignore email header information. Any line starting with \textcolor{blue}{---} is also ignored to ignore the separator in replies and forwards. Any word containing \textcolor{blue}{@} is also ignored to ignore email addresses. After which dot is added to the list of delimiters to account for the end of sentences but not the dot in email addresses. As a result 19401 words are selected which in this paper after being sorted lexicographically are called the initial parsed words and denoted by $W=(w_1,w_1,...,w_{19401})$.
\subsubsection{Counting the words and calculating normalized frequencies}
For each word $w_i \in W$ the number of emails containing that word is counted. Also for each word $w_i$ in each email $e_j$, the number of occurrences of that word in that email is counted and is normalized by dividing it over aggregate sum of occurrences of all words of $W$ in that email which the result is referred to as normalized per email word frequency in this paper and denoted by $T_{j_i}$.
\subsubsection{Filtering stop words}
The list of lowercase stop words in Table~\ref{stop_words} are used to filter the list of words in $W$ into a filtered list of words $\widehat{W}$.\\
\begin{table}
\begin{tabular}{ | c | p{12cm} |}
\hline
 Category 1 & i, we, you, they, she, he, it, this, that, these, those\\
\hline
 Category 2 & my, our, your, their, her, his, its\\
\hline
 Category 3 & me, us, you, them, her, him\\
\hline
 Category 4 & was, were, am, are, is, be, been, being, will, would, could, can, had, has, have, may, might, should\\
\hline
 Category 5 & and, or, but, also, however, so, because, not, if, then\\
\hline
 Category 6 & a, an, the\\
\hline
 Category 7 & what, who, which, where, when, how, did, do, does, why\\
\hline
 Category 8 & only, all, just, any, few, some, other, much, very\\
\hline
 Category 9 & one, two, three, four, five, six, seven, eight, nine, ten\\
\hline
 Category 10 & get, let, want, like\\
\hline
 Category 11 & here, there, in, of, for, at, as, with, by, on\\
\hline
 Category 12 & http, https, www, com\\
\hline
 Category 13 & to, from, subject, cc, bcc, re, fw, forwarded, sent, am, pm, attached, regards, best, find, email, following, thanks, thank, dear, hi, hello, fax, phone, address, e-mail, below, fyi\\
\hline
 Category 14 & eol\\
\hline
 Category 15 & monday, tuesday, wednesday, thursday, friday, saturday, sunday\\
\hline
 Category 16 & january, february, march, april, may, june, july, august, september, october, november, december\\
\hline
\end{tabular}
\caption{Stop words (lowercase)}
\label{stop_words}
\end{table}

\FloatBarrier

$\widehat{W}$ has 19256 words in it.

\subsubsection{Lexical filter}
From $\widehat{W}$, any single character word is filtered in addition to any word containing any digit or hypen or \textcolor{blue}{@} or \textcolor{blue}{\&} symbol. The result is called the basic filtered words in this paper and denoted by $\ddot{W}$. The number of basic filtered words is 13688.
\subsection{Finding the top indicative words for each email}
\subsubsection{Calculate per email significance for each of basic filtered words}
Term Frequency–Inverse Document Frequency~\cite{2_text_classification,3_beyond} is computed for each of basic filtered words per email and the result is called per email significance for each of basic filtered words in this paper and denoted by $S_{j_i}$ for each email $e_j \in E$ and each $\ddot{w_i} \in \ddot{w_i}$. Mathematically speaking:
$$S_{j_i}=\frac{F_{j_i}}{\sum_t{F_{a_t}}} \cdot\log(\frac{N}{\widehat{F}_i})$$ where $F_{j_i}$ is the number of times that the word $\ddot{w_i}$ appears in the email $e_j$, and $\widehat{F}_i$ specifies the number of emails that contain the word $\ddot{w_i}$ at least once, and $N$ is the number of selcted emails ($N=2400$).
\subsubsection{Calculate per email significance rank for each of basic filtered words}
Based on the per email significance for each of basic filtered words, the ranking of each of the basic filtered words in each of the emails is calculated.
\subsubsection{Selecting first frequency filtered words}
Out of 13688 basic filtered words only 2413 appear in at least 10 of the selected emails which those words are called the first frequency filtered words in this paper and denoted by $H$.
\subsubsection{Selecting second frequency filtered words}
Out of 2413 first frequency filtered words, only 427 appear in the top 100 words in at least 51 emails ranked based on the ranking of basic filtered words for that email and those 427 words after being lexicographically sorted are called second frequency filtered words in this paper and denoted by $\ddot{H}=(\ddot{h}_1,\ddot{h}_2,...,\ddot{h}_{427})$.
\subsubsection{Extracting per email significance and significance ranks for each of second frequency filtered words}
At this point, the focus is shifted from basic filtered words to second frequency filtered words, so only per email significance of those words are kept but the ranking per email is set to -1 if the word does not appear in the top 100 words for that email based on the ranking of the basic filtered words.
\subsection{Computing Cosine tailored features per email}
At this stage, the average number of times a word in the second frequency filtered words has appeared in the top 100 words of an email is computed and truncated to the integer part of that number. Then two times that average number is selected as the desired number of ones in the final features per email. Now out of 427 second frequency filtered words, when they appear in the top 100 list for an email (except if the ranking is -1), the feature for that word for that email is considered one. The issue however is that the number of ones is not fixed for each email as desired , so for each email, enough ones are added to make the number of ones equal for all emails but a dimensions used for a dummy 1 for an email cannot be used for a dummy 1 for another email, which leads to significant number of features. The total number of features per email at the end is 48557.

\subsection{Scrambling the sorting of selected emails}
At the end, before creating the final data set and before evaluation,  the order of emails is scrambled using the pseudo-random number generator of go programming language (rand.New and rand.Rand.shuffle functions) initialized with a seed value. Although not perfect, the pseudo random number generator should be practically good enough for removing any tangible bias based on the order of emails in most or all cases (The author cannot imagine a case where more sophisticated random number generators such as the CSPRNG (Cryptographically Secure Pseudorandom Number Generator) type of random generators are needed for this task but the extent of his imagination does not necessarily cover all possible cases).

\section{Evaluation and statistics}
\label{evaluation_and_statistics}
The evaluation is based on the accuracy of a $k$ nearest neighbours majority voting classification using Cosine distance. In addition to KNN majority voting classification accuracy and confusion matrix, some statistics for the process is reported. The KNN majority voting classification accuracy using Cosine distance with $k=10$ is 76.75\% which shows at least some level of success given the 8 labels involved. The result of conversion is 48557 features per selected email out of which exactly 40 features per email are non-zero. Average number of non zero primary features per email is 19.946 (rounded to 3 decimal places) . Standard deviation of number of non zero primary features per email: 10.963 (rounded to 3 decimal places). The table~\ref{per_label} shows average number of non zero primary features per email per label number (rounded to 3 decimal places) and standard deviation of number of non zero primary features per email per label number (rounded to 3 decimal places).

\begin{table}
\begin{tabular} { | c | c | c |}
\hline
Label number & Average & Standard deviation \\
\hline
0 &  21.407 & 9.706 \\
\hline
1 &  17.190 & 10.510 \\
\hline
2 &  26.230 & 10.639 \\
\hline
3 &  27.177 & 10.009 \\
\hline
4 &  11.517 & 8.413 \\
\hline
5 &  13.857 & 2.026 \\
\hline
6 &  22.430 & 10.954 \\
\hline
7 &  19.760 & 11.522 \\
\hline
\end{tabular}
\caption{average number of non zero primary features per email per label number (rounded to 3 decimal places) and standard deviation of number of non zero primary features per email per label number (rounded to 3 decimal places)}
\label{per_label}
\end{table}

\begin{figure}
\includegraphics[width=14cm]{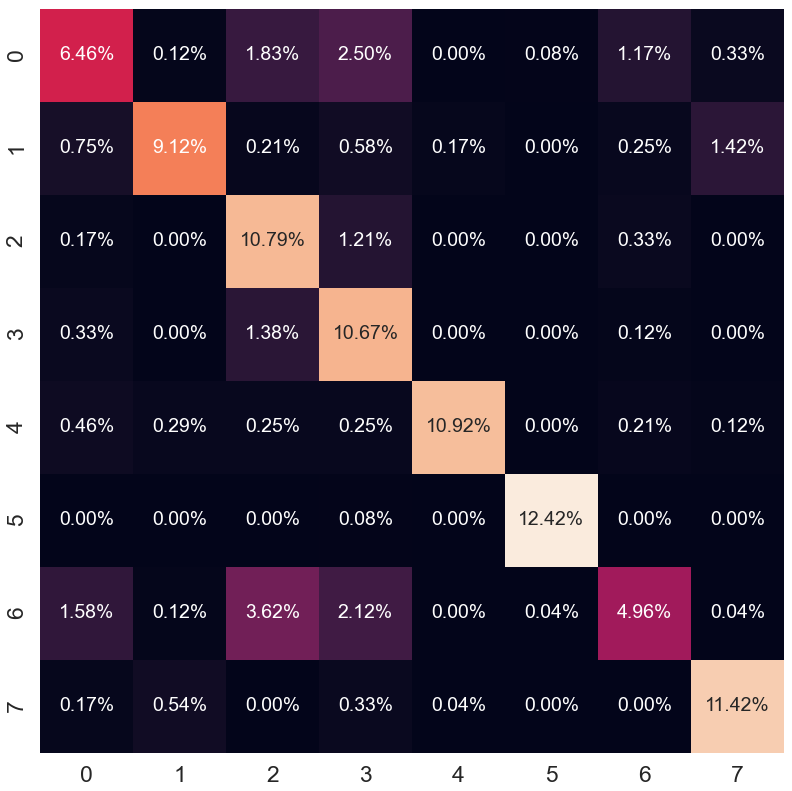}
\caption{ Heatmap of the confusion matrix (rounded to two decimal places, the rounding is also reflected in colours) of the KNN majority voting classification}
\label{confusion_matrix_heatmap}
\end{figure}

The Table~\ref{confusion_matrix_table} and Figure~\ref{confusion_matrix_heatmap} show the confusion matrix for the KNN majority voting classification.

\begin{table}
\begin{tabular}{ | c | c | c | c | c | c | c | c | c | }
\hline
&0&1&2&3&4&5&6&7\\
\hline
0 & 6.46\% & 0.12\% & 1.83\% & 2.50\% & 0.00\% & 0.08\% & 1.17\% & 0.33\% \\
\hline
1 & 0.75\% & 9.12\% & 0.21\% & 0.58\% & 0.17\% & 0.00\% & 0.25\% & 1.42\% \\
\hline
2 & 0.17\% & 0.00\% & 10.79\% & 1.21\% & 0.00\% & 0.00\% & 0.33\% & 0.00\% \\
\hline
3 & 0.33\% & 0.00\% & 1.38\% & 10.67\% & 0.00\% & 0.00\% & 0.12\% & 0.00\% \\
\hline
4 & 0.46\% & 0.29\% & 0.25\% & 0.25\% & 10.92\% & 0.00\% & 0.21\% & 0.12\% \\
\hline
5 & 0.00\% & 0.00\% & 0.00\% & 0.08\% & 0.00\% & 12.42\% & 0.00\% & 0.00\% \\
\hline
6 & 1.58\% & 0.12\% & 3.62\% & 2.12\% & 0.00\% & 0.04\% & 4.96\% & 0.04\% \\
\hline
7 & 0.17\% & 0.54\% & 0.00\% & 0.33\% & 0.04\% & 0.00\% & 0.00\% & 11.42\% \\
\hline
\end{tabular}
\caption{Confusion matrix (rounded to two decimal places) of the KNN majority voting classification}
\label{confusion_matrix_table}
\end{table}

The table~\ref{some_of_the_statistics} shows some of the statistics related to the process of creating features.

\begin{table}
\begin{tabular}{| p{12cm} | c |}
\hline
Number of initial parsed words& 19401\\
\hline
Number of stop words filtered words& 19401\\
\hline
Number of basic filtered words& 13688\\
\hline
Number of first frequency filtered words& 2413\\
\hline
Number of second frequency filtered words& 427\\
\hline
Average second frequency filtered words occurrence in per email top rankings for basic filtered words (truncated to integer part)& 20\\
\hline
Number of non zero primary features per email upper bound& 40\\
\hline
Number of features per email& 48557\\
\hline
Number of primary features per email& 427\\
\hline
Number of secondary features per email& 48130\\
\hline
Average number of secondary non zero features per email (rounded to 3 decimal places)& 20.054\\
\hline
\end{tabular}
\caption{Some of the statistics}
\label{some_of_the_statistics}
\end{table}

\FloatBarrier

\section{Final data set}
\label{final_data_set}
The result of conversion is a data set named MeeeFTCD (Massive Enhanced Extracted Email Features Tailored for Cosine Distance) which is available at \url{https://github.com/farshad-barahimi-academic-codes/data_sets_preparation/blob/main/data_sets/meeeftcd_data_set.zip} (access date: May 10, 2022) and \url{https://web.cs.dal.ca/~barahimi/data-sets/meeeftcd/} . The code used to create the data set and the results of this paper can be found at \url{https://web.cs.dal.ca/~barahimi/data-sets/meeeftcd/code/} (access date: May 11,2022) for which a guide is available at \url{https://web.cs.dal.ca/~barahimi/data-sets/meeeftcd/how-was-built/} (access date: May 11,2022) which relies on compiling the code written in go programming language and passing right arguments to it.

\bibliography{preprint}

\end{document}